\documentclass[preprint,1p]{elsarticle}

\usepackage{graphicx,xcolor,amsmath,amssymb,pifont,epstopdf}
\journal{Journal of \LaTeX\ Templates}

\begin{document}
\begin{frontmatter}

\title{Chimera states in nonlocally coupled bicomponent phase oscillators: From synchronous to asynchronous chimeras}

\author[rvt]{Qionglin Dai}

\author[rvt]{Kai Yang}

\author[rvt]{Hongyan Cheng}

\author[rvt]{Haihong Li}

\author[rvt2]{Fagen Xie\corref{mycorrespondingauthor}}
\ead{xiefagen@yahoo.com}

\author[rvt]{Junzhong Yang\corref{mycorrespondingauthor}}
\cortext[mycorrespondingauthor]{Corresponding author}
\ead{jzyang@bupt.edu.cn}

\address[rvt]{School of Science, Beijing University of Posts and Telecommunications,\\ Beijing, 100876, People's Republic of China}
\address[rvt2]{Department of Research and Evaluation, Kaiser Permanente Southern California, Pasadena, CA 91101, USA}

\begin{abstract}
Chimera states, a symmetry-breaking spatiotemporal pattern in nonlocally coupled identical dynamical units, prevail in a variety of systems. Here, we consider a population of nonlocally coupled bicomponent phase oscillators in which oscillators with natural frequency $\omega_0$ (positive oscillators) and $-\omega_0$ (negative oscillators) are randomly distributed along a ring. We show the existence of chimera states no matter how large $\omega_0$ is and the states manifest themselves in the form that oscillators with positive/negative frequency support their own chimera states. There are two types of chimera states, synchronous chimera states at small $\omega_0$ in which coherent positive and negative oscillators share a same mean phase velocity and asynchronous chimera states at large $\omega_0$ in which coherent positive and negative oscillators have different mean phase velocities. Increasing $\omega_0$ induces a desynchronization transition between synchronous chimera states and asynchronous chimera states.
\end{abstract}

\end{frontmatter}

%\linenumbers

\section{Introduction}
In the past decade, we have witnessed rapid expansion of the field of chimera states when the states evolve from a surprising symmetry-breaking spatiotemporal pattern to a prevailing dynamical phenomenon ranging from physics and chemistry to biology and from classical to quantum systems \cite{kura,abra,lai09,mot10,zhu12,pan15,mart13,tins12,hage12,cheng18,gav18,lai18}. Chimera states, characterized by the alternation between coherent and incoherent regions, were originally found in nonlocally coupled identical phase oscillators \cite{kura}. Later, chimera states have been observed in periodic and chaotic maps \cite{omel11}, mechanical oscillators \cite{omel15}, neuronal oscillators \cite{omel13,hiz14,sak06}, and chemical oscillators \cite{tot18}. Two ingredients are thought to be required for chimera states, self-oscillating units and the nonlocal coupling. However, the two requirements have been relaxed recently. Chimera states have been found in global \cite{yel14,set14,cha14,pre15} as well as in local interactions \cite{lai15,ber16}. Furthermore, excitable systems, ubiquitous in biology, chemistry, and physics, allow for a stable equilibrium and, responding to strong perturbation, they go back to the equilibrium only after a large excursion. Chimera states have been studied in nonlocally coupled excitable systems in the presence of noise \cite{sem16}. Dai et al. found that chimera states emerge out of excitable units through a coupling-induced collective oscillation \cite{dai18}.

For now, due to the prevalence of chimera states in various systems, the investigation on chimera states may have been shifted from finding chimera states in various systems to exploring the connection between chimera states and other dynamical phenomena. Motter et al. proposed a connection between chimera states and cluster synchronization in networks of locally coupled chaotic oscillators \cite{mot17}. Lai et al. established a connection between chimera states and a quantum scattering phenomenon in 2-dimensional Dirac material systems where manifestations of classically integrable and chaotic dynamics coexist simultaneously \cite{lai18}. In this work, we study a ring of nonlocally coupled bicomponent phase oscillators and report an interesting transition from synchronized to desynchronized chimera states.

The rest of paper is organized as follows. In Section 2, we present the model of nonlocally coupled bicomponent phase oscillators in a ring and give a brief summary on chimera states in a ring of identical phase oscillators. In Section 3, we demonstrate the existence of chimera states in the bicomponent phase oscillators and a desynchronization transition induced by the frequency mismatch among oscillators. We also present a theoretical analysis on the observed chimera states in this section. Finally, we conclude with a summary in Section 4.

\section{Model}
As a standard model exploring chimera states, a ring of nonlocally coupled phase oscillators is described as
\begin{eqnarray}\label{eq1}
\dot{\theta}_{i}(t)&=&\omega-\sum_{\text{j=1}}^{\text{N}}G(\vert i-j\vert)\sin[\theta_{i}(t)-\theta_{j}(t)+\alpha].
\end{eqnarray}
The subscript $i$ refers to the node index, which has to be taken
modulo $N$ (or period boundary condition). $\theta_{i}$ represents the phase of oscillator $i$. $\omega$ and $\alpha$ are the natural frequency of oscillators and the phase lag, respectively. The kernel function $G(\vert i-j\vert)$, the nonlocal coupling between oscillators $i$ and $j$, is assumed to be even, nonnegative, decreasing with $|i-j|$ and normalized to have unit summation. Defining position-dependent complex order parameters
$Z_{i}=R_{i}e^{i\Theta_{i}}=\sum_{\text{j=1}}^{\text{N}}G(\vert i-j\vert)e^{i\theta_{j}}$, Eq.~(\ref{eq1}) is reformulated as
\begin{eqnarray}\label{eq2}
\dot{\phi}_{i}&=&\omega-\dot{\Theta}_{i}-R_{i}\sin(\phi_{i}+\alpha)
\end{eqnarray}
by letting $\phi_{i}=\theta_{i}-\Theta_{i}$.

For kernel function $G(x)=(1+A\cos2\pi x/N)/2\pi$ with $0\leq A\leq1$, Eq.~(\ref{eq1}) allows for both complete synchronous state and chimera states for proper $A$ and $\alpha$. When the heterogeneity in natural frequency is introduced, Laing found that chimera states are robust as the dispersion of natural frequencies is not sufficiently strong \cite{lai09}. Previous studies have shown some common features of chimera states in Eq.~(\ref{eq1}) with sufficient large $N$ as following: $\dot{\Theta}_{i}$ is a constant independent of position and synchronized oscillators are frequency-locked to $\Theta_i$ when $|\omega-\dot{\Theta}_{i}|<R_{i}$.

Here, we consider a simplest heterogeneity in the natural frequencies of oscillators by assuming that the frequency distribution follows a double-delta function $g(\omega)=p\delta(\omega-\omega_0)+(1-p)\delta(\omega+\omega_0)$ with $p\in(0,1)$. Every oscillator is assigned with a natural frequency $\omega_0$ with the probability $p$ and $-\omega_0$ with the probability $1-p$. For convenience, we term oscillators with positive/negative natural frequency as positive/negative oscillators. The heterogeneity in natural frequency is measured by $\omega_0$. We are interested in how the heterogeneity impacts on chimera dynamics. Considering a coupled system with two different phase oscillators, we know that, depending on the frequency mismatch between oscillators, the coupled oscillators may display a transition between synchronization and desynchronization. We are interested in whether such a transition occurs in a ring of nonlocally coupled bicomponent phase oscillators.

\section{Simulation results and analysis}
We start with $p=0.5$, $A=1$ and $\alpha=1.45$ and numerically simulate Eq.~(\ref{eq1}) using a fourth-order Runge-Kutta algorithm with $\delta t=0.025$. The number of oscillators $N=256$. Initially, each oscillator randomly takes its phase in the range of $[0,2\pi]$.

\begin{figure*}
\centering
\includegraphics[width=3.5in]{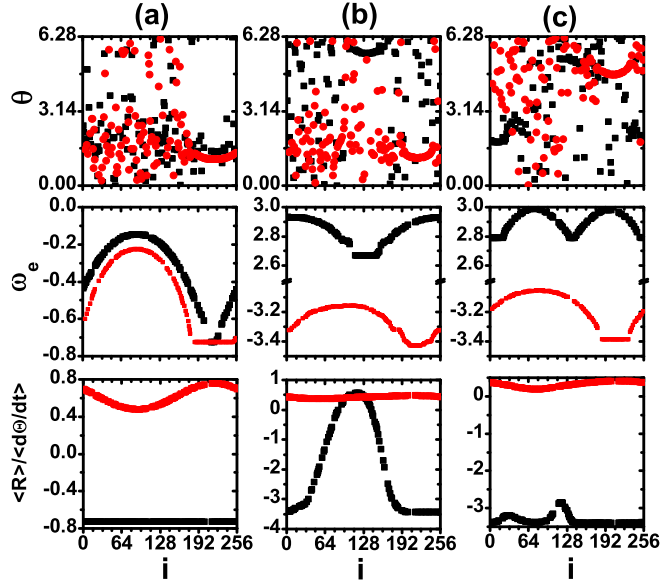}
\caption{\label{fig1}The chimera states at $\omega_0=0.03$ (a), $\omega_0=3$ (b) and (c). The top row shows the snapshots of $\theta$ for negative oscillators (red) and positive oscillators (black). The second row shows the mean phase velocities $\omega_e$ which condensates onto two curves in corresponding with negative oscillators (red) and positive oscillators (black). The third row shows $\langle R\rangle$ (red) and $\langle d\Theta/dt\rangle$ (black). $A=1$ and $\alpha=1.45$.}
\end{figure*}

The top row in Fig.~\ref{fig1} shows the snapshots of oscillators for $\omega_0=0.03$ and $\omega_0=3$. When we take all oscillators in a whole, we find one small coherent cluster for $\omega_0=0.03$ while no coherent cluster for $\omega_0=3$. However, if we take positive and negative oscillators separately, we find that each group of oscillators supports their own chimera states. At $\omega=0.03$, the coherent cluster for positive oscillators (the black dots) is smaller than that for negative oscillators (the red dots) and the two coherent clusters are overlapped in space. At $\omega_0=3$, the coherent clusters for positive and negative oscillators are not overlapped, which prevent the chimera states from direct observation if we do not distinguish positive and negative oscillators. There are two types of chimera dynamics for positive oscillators at $\omega_0=3$, chimera state with one coherent cluster in Fig.~\ref{fig1}(b) and chimera state with two coherent clusters in Fig.~\ref{fig1}(c), in contrast to that only one coherent cluster exists for negative oscillators.

\begin{figure*}
\includegraphics[width=6in]{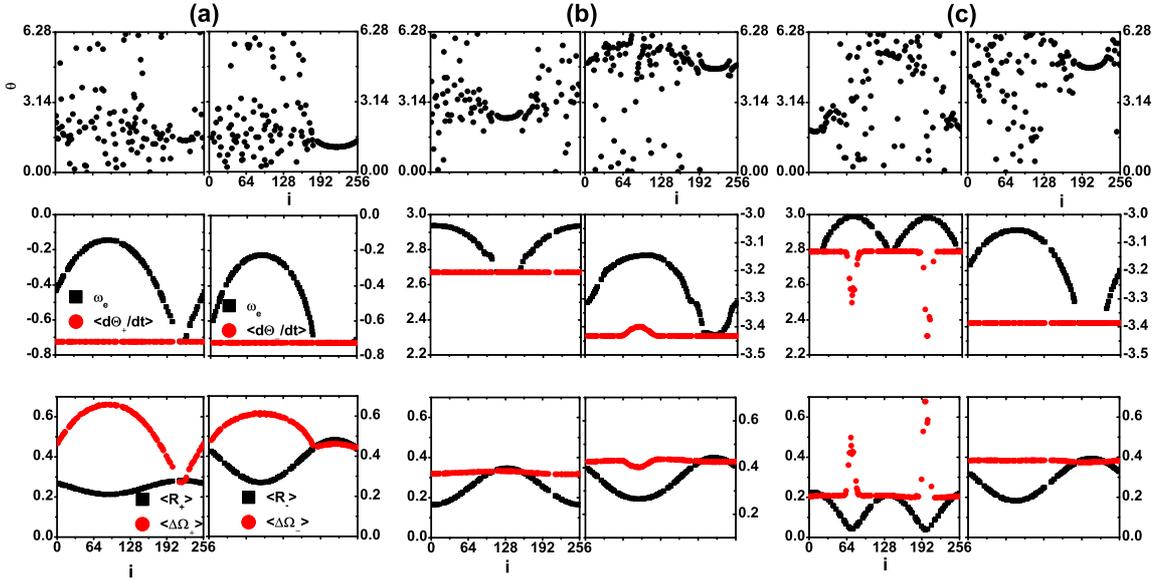}
\caption{\label{fig2}The chimera states at $\omega_0=0.03$ (a), $\omega_0=3$ (b) and (c). In each plot, the left column shows the results for positive oscillators and the right column for negative oscillators. The top row shows the snapshots of $\theta$, the middle row shows $\omega_e$ (black) and $\langle d\Theta_{\pm}/dt\rangle$ (red), and the bottom row shows $\langle R_{\pm}\rangle$ (black) and $\langle \Delta\Omega_{\pm}\rangle$ (red). $A=1$ and $\alpha=1.45$.}
\end{figure*}

The mean phase velocity of an oscillator, defined as $\omega_{e,i}=\langle\dot{\theta}_{i}\rangle$ with $\langle\cdot\rangle$ the time average over a long time interval, is often used to distinguish the coherent and incoherent oscillators. The profiles of $\omega_{e,i}$ in the second row in Fig.~\ref{fig1} show that positive oscillators and negative oscillators condensate onto two different curves. At $\omega_0=0.03$, the two curves share the same mean phase velocity for oscillators in the coherent clusters while the mean phase velocities in the coherent clusters for positive and negative oscillators are well separated at $\omega_0=3$. Furthermore, we monitor the spatially-dependent complex order parameter $Z_i$. $\langle \dot{\Theta}_{i}\rangle $ and $\langle R_i\rangle $ averaged over a long time interval are presented in the bottom row in Fig.~\ref{fig1}. $\langle \dot{\Theta}_{i}\rangle $ is independent of space at $\omega_0=0.03$ while it becomes patterned in space at $\omega_0=3$. Especially, in the chimera state with one coherent cluster for positive oscillators, the inhomogeneity in $\langle \dot{\Theta}_{i}\rangle $ ranging from -3.4 to 0.5 is much strong. Moreover, the criteria $|\omega_i-\langle\dot{\Theta}_{i}\rangle |<R_{i}$ for coherent clusters in chimera state fails for $\omega_0=3$. In short, the chimera states at $\omega_0=3$ can not be explained by following Eq.~(\ref{eq2}) and the global view not differentiating positive and negative oscillators fails in the description of chimera states in Figs.~\ref{fig1}(b) and (c).

It is quite surprising and fascinating that positive and negative oscillators may maintain their own chimera states when positive and negative oscillators are randomly mixed in space. However, the phenomena can be understood as follows. Bearing in mind that positive and negative oscillators support their own chimera states, we introduce complex order parameters for positive and negative oscillators, respectively,
\begin{eqnarray}\label{eq3}
Z_{\pm,i}=R_{\pm,i}e^{i\Theta_{\pm,i}}=\sum_{j\in S_{\pm}}G(\vert i-j\vert)e^{i\theta_{\pm,j}}
\end{eqnarray}
with $\pm$ representing positive and negative oscillators and $S_{\pm}$ the set of positive and negative oscillators. Then Eq.~(\ref{eq1}) is reformulated as
\begin{eqnarray}\label{eq4}
\dot{\phi}_{\pm,i}&=&\pm\omega_0-\dot{\Theta}_{\pm,i}-R_{\pm,i}\sin(\phi_{\pm,i}+\alpha)\nonumber\\
&&-R_{\mp,i}\sin(\phi_{\pm,i}+\Theta_{\pm,i}-\Theta_{\mp,i}+\alpha)
\end{eqnarray}
with $\phi_{\pm,i}=\theta_{\pm,i}-\Theta_{\pm,i}$. We conjecture that, in the chimera states in Fig.~\ref{fig1}, coherent positive and negative oscillators get trapped by $Z_{\pm}$, respectively. That is, defining the quantity $\Delta\Omega_{\pm,i}$ as
\begin{eqnarray}\label{eq5}
\Delta\Omega_{\pm,i}=\pm\omega_0-\dot{\Theta}_{\pm,i}-R_{\mp,i}\sin(\phi_{\pm,i}+\Theta_{\pm,i}-\Theta_{\mp,i}+\alpha),\nonumber
\end{eqnarray}
we conjecture that a coherent oscillator should satisfy the condition $|\langle\Delta\Omega_{\pm,i}\rangle|<\langle R_{\pm,i}\rangle$. In Fig.~\ref{fig2}, the positive and negative oscillators are plotted separately. From the middle row, we find that $\langle \dot{\Theta}_{\pm,i}\rangle $ are in coincidence with the mean phase velocities of coherent positive and negative oscillators, respectively. The bottom row in Fig.~\ref{fig2} shows  $|\langle\Delta\Omega_{\pm,i}\rangle|<\langle R_{\pm,i}\rangle$ only in coherent clusters, which demonstrates the validity of the description considering positive and negative oscillators separately.

There is one extraordinary feature in Fig.~\ref{fig2}, $\langle\dot{\Theta}_{+,i}\rangle=\langle\dot{\Theta}_{-,i}\rangle$ at $\omega_0=0.03$ while the equality is broken at $\omega_0=3$. The feature reveals the synchronization between the chimera dynamics for positive and negative oscillators at $\omega_0=0.03$ and the desynchronization between these two chimera dynamics at $\omega_0=3$. We term the former states as synchronous chimera states and the latter asynchronous chimera states. $\langle\dot{\Theta}_{\pm,i}\rangle$ stays at a constant in the synchronous chimera states but displays kind of inhomogeneity in space in the asynchronous chimera states.

\begin{figure}
  \centering
  \includegraphics[width=3in]{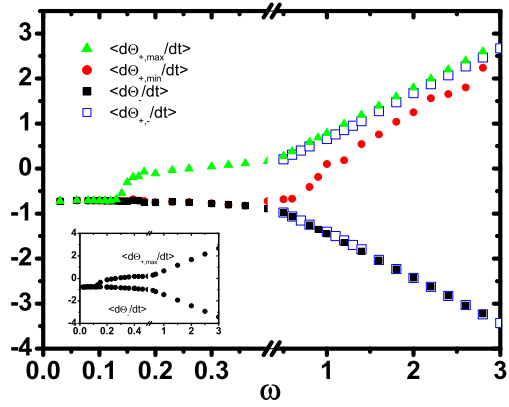}
  \caption{\label{fig3}The desynchronization transition from synchronous chimera states to asynchronous chimera states at $A=1$ and $\alpha=1.45$. The solid symbols are acquired from synchronous chimera states and asynchronous chimera states with two coherent clusters in positive oscillators and the open symbols are from the asynchronous chimera states with one coherent clusters in positive oscillators. The inset shows another example at $A=0.89$ and $\alpha=1.38$.}
\end{figure}

\begin{figure}
  \centering
  \includegraphics[width=3.5in]{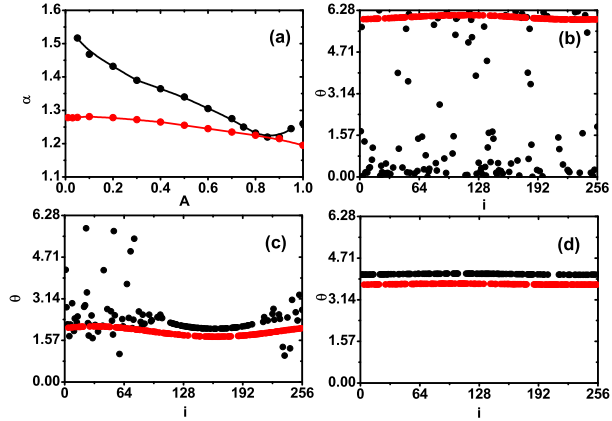}
  \caption{\label{fig4}(a) The stability diagrams in the plane of $A$ and $\alpha$ of synchronous chimera states at $\omega_0=0.06$ (red) and asynchronous chimera states at $\omega_0=3$ (black). (b-d) The dynamical states represented by the snapshots $\theta$ at $\alpha=1.31$, $\alpha=1.24$, and $\alpha=1.1$ show the transition scenario from synchronous chimera states to synchronous states at $\omega=0.06$ and $A=1$.}
\end{figure}

The transition between the synchronous and asynchronous chimera states can be investigated by monitoring $\langle\dot{\Theta}_{\pm,i}\rangle$. Since $\langle\dot{\Theta}_{+,i}\rangle$ exhibits strong inhomogeneity in asynchronous chimera states with two coherent clusters, we also record its maximum and its minimum for each $\omega_0$. As shown in Fig.~\ref{fig3}, the transition occurs at around $\omega_{0c}=0.12$ where the maximum of $\langle\dot{\Theta}_{+,i}\rangle$ departs from $\langle\dot{\Theta}_{-,i}\rangle$. Before $\omega_{0c}$, $\langle\dot{\Theta}_{\pm,i}\rangle$ is a constant and $\langle\dot{\Theta}_{+,i}\rangle=\langle\dot{\Theta}_{-,i}\rangle$ is held, which suggests the synchronous state between chimera dynamics for positive and negative oscillators. Beyond $\omega_{0c}$, the asynchronous chimera state with two coherent clusters first appears and the asynchronous chimera state with one coherent cluster steps in at around $\omega_0=0.4$. The inset in Fig.~\ref{fig3} shows the transition at other combination of $A$ and $\alpha$ where $\langle\dot{\Theta}_{-,i}\rangle$ and the maximum of $\langle\dot{\Theta}_{+,i}\rangle$ are plotted, which supports that the transition between synchronous and asynchronous chimera states for positive and negative oscillators is not exceptional. To be brief, Fig.~\ref{fig3} reports a desynchronization transition with $\omega_0$ between two chimera dynamics, in positive oscillators and in negative oscillators, simultaneously existing in a same population, which is kind of interesting observation in the field of chimera states.

The existence of chimera states in identical phase oscillators is dependent on the parameters  $A$ and $\alpha$. We are curious about how robust the synchronous chimera states and asynchronous chimera states are to the variation of parameters $A$ and $\alpha$ in bicomponent phase oscillators. We focus on synchronous chimera states at $\omega_0=0.06$ and asynchronous chimera states at $\omega_0=3$. For convenience, we will not distinguish the asynchronous chimera states with two coherent clusters and those with one coherent cluster. To acquire the boundary of the stability diagram for these chimera states, we consider two paths by continuously decreasing $A$ with fixed $\alpha$ or continuously decreasing $\alpha$ with fixed $A$. The stability diagrams are presented in Fig.~\ref{fig4}(a). We find that the critical $\alpha$ tends to increase with $A$ decrease for both types of chimera states and the parameter regime for synchronous chimera states is much larger than that for asynchronous chimera states. In comparison with the stability diagram of chimera state in nonlocally coupled identical phase oscillators \cite{abra}, a ring of bicomponent phase oscillators supports chimera states in much larger parameter regimes. Figures~\ref{fig4}(b-d) show a transition scenario at $\omega_0=0.06$ and $A=1$. With $\alpha$ decrease, positive oscillators first transit to incoherence and then back to chimera dynamics, and finally to an all-coherent state. On the other hand, chimera dynamics in negative oscillators is replaced by an all-coherent dynamics with $\alpha$ decrease. Combining these together, we may find incoherent-coherent state, chimera-coherent state, and coherent-coherent state in Fig.~\ref{fig4}(b-d), respectively.

\begin{figure}
  \centering
  \includegraphics[width=3.5in]{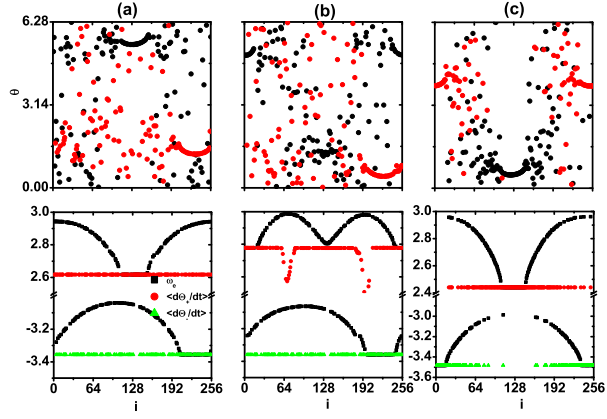}
  \caption{\label{fig5}(a) and (b) Chimera states with one coherent cluster in positive oscillators and with two coherent clusters in positive oscillators at $p=0.6$. (c) Chimera states with one coherent cluster in positive oscillators with $p(i)=\sin(\pi i/N)$. The top panels show the snapshots for positive oscillators (black) and negative oscillators (red). The bottom panels show the mean phase velocity (black), $\langle\dot{\Theta}_{+,i}\rangle$ (red) and $\langle\dot{\Theta}_{-,i}\rangle$ (green). $A=1$, $\alpha=1.45$, and $\omega_0=3$.}
\end{figure}

The parameter $p$ accounts for the fraction of positive oscillators in the population of phase oscillators. The qualitative properties of the model dynamics are not sensitive to the value of $p$. In Figs.~\ref{fig5}(a) and (b), we show an example at $p=0.6$ where asynchronous chimera states with one coherent cluster and with two coherent clusters at $\omega_0=3$ are presented, respectively. Furthermore, the model dynamics keeps unchanged even when $p$ is position-dependent. The synchronous chimera states are realized when the maximum of position-dependent $p$ is small and asynchronous chimera states otherwise. In Fig.~\ref{fig5}(c), we show the asynchronous chimera states with one-coherent cluster at $\omega_0=3$ for $p(i)=\sin(\pi i/N),i=1,2,\cdots,N$.

\begin{figure}
  \centering
  \includegraphics[width=3.5in]{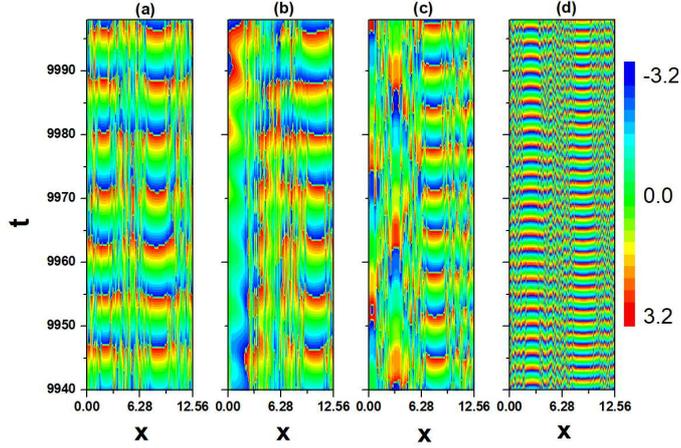}
  \caption{\label{fig6}The evolutions of the arguments of $a_{\pm}$ acquired by simulating Eq.~(\ref{eq9}) at $\omega_0=0.03$ in (a), $\omega_0=0.2$ in (b), $\omega_0=0.5$ in (c), $\omega_0=3$ in (d). For the better illustration, we shift the space variable of $a_{-}$ from $x$ to $x+2\pi$. $A=1$, $\alpha=1.45$, and $p=0.5$.}
\end{figure}

\begin{figure}
  \centering
  \includegraphics[width=3.5in]{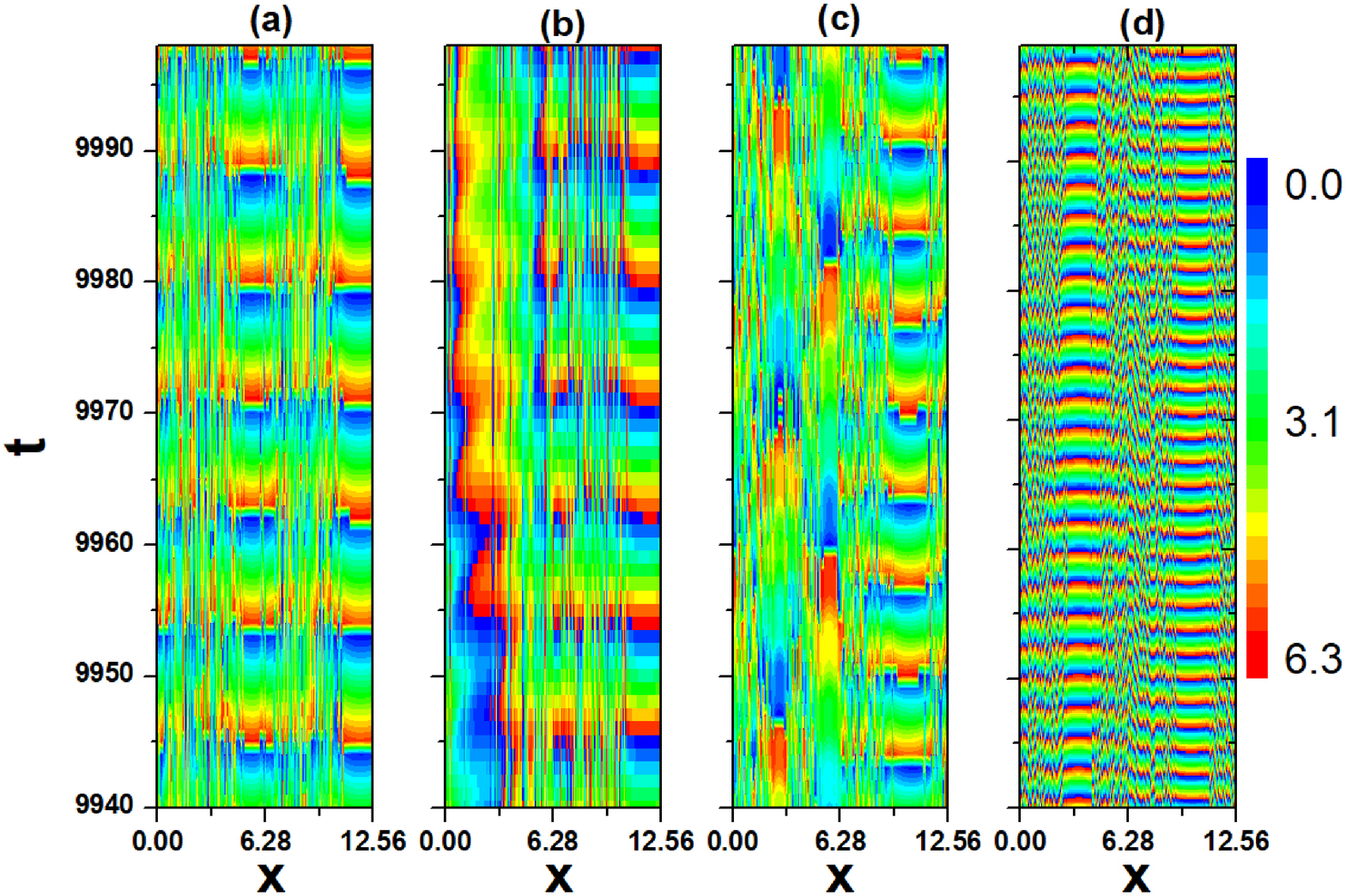}
  \caption{\label{fig7}The evolutions of $\theta_{\pm}$ in Eq.~(1) where positive and negative oscillators alternate along the ring at $\omega_0=0.03$ in (a), $\omega_0=0.2$ in (b), $\omega_0=0.5$ in (c), $\omega_0=3$ in (d). For the better illustration, we shift the space variable of negative oscillators from $x$ to $x+2\pi$. $A=1$, $\alpha=1.45$, and $p=0.5$.
}
\end{figure}

The ring of nonlocally coupled bicomponent phase oscillators, when the number of phase oscillators goes to infinity, can be analyzed with the assistance of the Ott-Antonsen ansatz \cite{lai09,ott08}. We set the length of the ring to be $2\pi$ and assume that there are probability density functions $f_{\pm}(x,\omega,\theta,t)$ characterising the states of the positive and negative oscillators, respectively. The two functions satisfy the continuity equations
\begin{eqnarray}\label{eq5}
\frac{\partial f_{\pm}}{\partial t}+\frac{\partial(f_{\pm}v_{\pm})}{\partial \theta}=0.
\end{eqnarray}
Reformulating the order parameters for positive and negative oscillators
\begin{eqnarray}\label{eq6}
Z_{\pm}(x)=\int_{0}^{2\pi}G(x-x')\int_{0}^{2\pi}e^{i\theta}f_{\pm}d\theta dx',
\end{eqnarray}
we have
\begin{eqnarray}\label{eq7}
v_{\pm}=\omega_{0,\pm}-\frac{1}{2i}[Z^*e^{i(\theta+\alpha)}-Ze^{-i(\theta+\alpha)}]
\end{eqnarray}
where $Z^{*}$ denotes the complex conjugate of $Z$ and $Z=pZ_{+}+(1-p)Z_{-}$. Using Ott-Antonsen ansatz, we write the probability density functions as
\begin{eqnarray}\label{eq8}
f_{\pm}(x,\omega,\theta,t)=\frac{1}{2\pi}\{1+\sum_{n=1}^{\infty}[a_{\pm}^n(x,t)e^{in\theta}+c.c.]\}
\end{eqnarray}
where $c.c.$ is the complex conjugate of the previous term. By substituting Eqs.~(\ref{eq7}) and ~(\ref{eq8}) into Eqs.~(\ref{eq5}) and ~(\ref{eq6}), we obtain
\begin{eqnarray}\label{eq9}
\frac{\partial a_{\pm}(x,t)}{\partial t}&=&- i\omega_{0,\pm}+\frac{1}{2}[Z^*(x,t)e^{i\alpha}-Z(x,t)e^{-i\alpha}a_{\pm}^2],\nonumber\\
Z(x,t)&=&\int_{0}^{2\pi}G(x-x')[pa^*_{+}(x')+(1-p)a^*_{-}(x')]dx'.
\end{eqnarray}
Numerically simulating Eq.~(\ref{eq9}), we have the evolutions of $a_{\pm}(x,t)$. We present the arguments of $a_{\pm}(x,t)$ in Fig.~\ref{fig6} for several $\omega_0$ by shifting space variable of $a_{-}(x,t)$ to $x+2\pi$. In contrary to Fig.~\ref{fig1} and  Fig.~\ref{fig2}, the coherent clusters in $a_{+}$ and $a_{-}$ at $\omega_0=3$ are overlapped in space. To get the same chimera dynamics in a finite number of phase oscillators as those in Fig.~\ref{fig6}, we should arrange bicomponent phase oscillators with the alternation of positive and negative oscillators along the ring. Following the arrangement, the chimera dynamics acquired from the model Eq.~\ref{eq1} is presented in Fig.~\ref{fig7} which looks exactly same as those in Fig.~\ref{fig6}.

\section{Conclusion}
In conclusion, we have studied a ring of nonlocally coupled bicomponent phase oscillators where oscillators are randomly assigned with natural frequency $\omega_0$ and $-\omega_0$. We find the existence of chimera states no matter how large $\omega_0$ is. Different from ordinary chimera states in nonlocally coupled identical phase oscillators, the existence of chimera states may be concealed if improper measurement is taken. By considering positive and negative oscillators separately, we find that positive and negative oscillators support their own chimera dynamics. There are two types of chimera states in a ring of bicomponent phase oscillators, synchronous chimera states at small $\omega_0$ where the mean phase velocities in the coherent clusters for positive and negative oscillators are the same and asynchronous chimera states at large $\omega_0$ where coherent positive and coherent negative oscillators have different mean phase velocities. Increasing $\omega_0$ may induce the desynchronization transition from synchronous to asynchronous chimera states. The synchronous and asynchronous chimera states may exist in an infinite number of oscillators, which is supported by the theoretical analysis based on Ott-Antonsen ansatz.

\section*{Acknowledgements}
This work is supported by National Natural Science Foundation of China under Grant Nos. 11575036 and 11505016.

\section*{References}
%\expandafter\ifx\csname url\endcsname\relax
 % \def\url#1{\texttt{#1}}\fi
%\expandafter\ifx\csname urlprefix\endcsname\relax\def\urlprefix{URL }\fi
%\expandafter\ifx\csname href\endcsname\relax
%  \def\href#1#2{#2} \def\path#1{#1}\fi


\begin{thebibliography}{}

\bibitem{kura} Kuramoto Y, Battogtokh D. Coexistence of coherence and incoherence in nonlocally coupled phase oscillators. Nonlinear Phenom Complex Syst 2002;5:380-5.

\bibitem{abra} Abrams DM, Strogatz SH. Chimera states for coupled oscillators. Phys Rev Lett 2004;93:174102.

\bibitem{lai09} Laing CR. The dynamics of chimera states in heterogeneous Kuramoto networks. Physica D 2009;238:1569-88.


\bibitem{mot10} Motter AE. Nonlinear dynamics: spontaneous synchrony breaking. Nat Phys 2010;6:164-5.

\bibitem{zhu12} Zhu Y, Li Y, Zhang M, Yang J. The oscillating two-cluster chimera state in non-locally coupled phase oscillators. EPL 2012;97:10009.

\bibitem{pan15} Panaggio MJ, Abrams DM. Chimera states: coexistence of coherence and incoherence in networks of coupled oscillators. Nonlinearity 2015;28:R67.

\bibitem{mart13} Martens EA, Thutupalli S, Fourri\`{e}re A, Hallatschek O. Chimera states in mechanical oscillator networks. Proc Natl Acad Sci USA 2013;110:10563-7.

\bibitem{tins12} Tinsley MR, Nkomo S, Showalter K. Chimera and phase-cluster states in populations of coupled chemical oscillators. Nat Phys 2012;8:662-5.

\bibitem{hage12} Hagerstrom AM, Murphy TE, Roy R, H\"{o}vel P, Omelchenko I, Sch\"{o}ll E. Experimental observation of chimeras in coupled-map lattices. Nat Phys 2012;8:658-61.

\bibitem{cheng18} Cheng H, Dai Q, Wu N, Feng Y, Li H, Yang J. Chimera states in nonlocally coupled phase oscillators with biharmonic interaction. Commun Nonlinear Sci Numer Simul 2018;56:1-8.

\bibitem{gav18} Gavrilov SS. Polariton chimeras: Bose-Einstein condensates with intrinsic chaoticity and spontaneous long-range ordering. Phys Rev Lett 2018;120:033901.

\bibitem{lai18} Xu H, Wang G, Huang L, Lai Y. Chaos in Dirac electron optics: emergence of a relativistic quantum chimera. Phys Rev Lett 2018;120:124101.

\bibitem{omel11} Omelchenko I, Maistrenko Y, H\"{o}vel P, Sch\"{o}ll E. Loss of coherence in dynamical networks: spatial chaos and chimera states. Phys Rev Lett 2011;106:234102.

\bibitem{omel15} Omelchenko I, Zakharova A, H\"{o}vel P, Siebert J, Sch\"{o}ll E. Nonlinearity of local dynamics promotes multi-chimeras. Chaos 2015;25:083104.

\bibitem{omel13} Omelchenko I, Omelchenko OE, H\"{o}vel P, Sch\"{o}ll E. When nonlocal coupling between oscillators becomes stronger: patched synchrony or multichimera states. Phys Rev Lett 2013;110:224101.

\bibitem{hiz14} Hizanidis J, Kanas V, Bezerianos A, Bountis T. Chimera states in networks of nonlocally coupled Hindmarsh-Rose neuron models. Int J Bifurcat Chaos 2014;24:1450030.

\bibitem{sak06} Sakaguchi H. Instability of synchronized motion in nonlocally coupled neural oscillators. Phys Rev E 2006;73:031907.

\bibitem{tot18} Totz JF, Rode J, Tinsley MR, Showalter K, Engel H. Spiral wave chimera states in large populations of coupled chemical oscillators. Nat Phys 2018;14:282-5.

\bibitem{yel14} Yeldesbay A, Pikovsky A, Rosenblum M. Chimeralike states in an ensemble of globally coupled oscillators. Phys Rev Lett 2014;112:144103.

\bibitem{set14} Sethia GC, Sen A. Chimera states: the existence criteria revisited. Phys Rev Lett 2014;112:144101.

\bibitem{cha14} Chandrasekar VK, Gopal R, Venkatesan A, Lakshmanan M. Mechanism for intensity-induced chimera states in globally coupled oscillators. Phys Rev E 2014;90:062913.

\bibitem{pre15} Premalatha K, Chandrasekar VK, Senthilvelan M, Lakshmanan M. Impact of symmetry breaking in networks of globally coupled oscillators. Phys Rev E 2015;91:052915.

\bibitem{lai15} Laing CR. Chimeras in networks with purely local coupling. Phys Rev E 2015;92:050904(R).

\bibitem{ber16} Bera BK, Ghosh D, Lakshmanan M. Chimera states in bursting neurons. Phys Rev E 2016;93:012205.

\bibitem{sem16} Semenova N, Zakharova A, Anishchenko V, Sch\"{o}ll E. Coherence-resonance chimeras in a network of excitable elements. Phys Rev Lett 2016;117:014102.

\bibitem{dai18} Dai Q, Zhang M, Cheng H, Li H, Xie F, Yang J. From collective oscillation to chimera state in a nonlocally coupled excitable system. Nonlinear Dyn 2018;91:1723-31.

\bibitem{mot17} Cho YS, Nishikawa T, Motter AE. Stable chimeras and independently synchronizable clusters. Phys Rev Lett 2017;119:084101.

\bibitem{ott08} Ott E, Antonsen TM. Low dimensional behavior of large systems of globally coupled oscillators. Chaos 2008;18:037113.

\end{thebibliography}
\end{document}